\documentclass[12pt,a4wide,epsf]{article}

\usepackage{epsfig}
\usepackage{graphicx}
\usepackage[german, english]{babel}
\usepackage{amsmath}
\usepackage{amssymb}
\usepackage{ifthen}
\usepackage{epsfig}

\newcounter{fig}   \newcommand{\lbfig}[1]{\refstepcounter{fig}
\label{#1} } 
\newcommand{\vphi}{\varphi}

\begin{document}

\title{
 Transitions between Vortex Rings\\
 and Monopole--Antimonopole Chains
\vspace{1.5truecm}
\author{
\vspace{0.5truecm}
{\bf Jutta Kunz, Ulrike Neemann, Yasha Shnir} \\
Institut f\"ur Physik, Universit\"at Oldenburg, Postfach 2503\\
D-26111 Oldenburg, Germany\\
}
}


\date{\today}

\maketitle
\vspace{1.0truecm}

\begin{abstract}
In monopole-antimonopole chain solutions of $SU(2)$ Yang-Mills-Higgs theory
the Higgs field vanishes at $m$ isolated points along the symmetry axis,
whereas in vortex ring solutions the Higgs field vanishes along
one or more rings, centered around the symmetry axis.
We investigate how these static axially symmetric solutions depend
on the strength of the Higgs selfcoupling $\lambda$.
We show, that as the coupling is getting large,
new branches of solutions appear
at critical values of $\lambda$.
Exhibiting a different node structure, these
give rise to transitions between
vortex rings and monopole-antimonopole chains.
\end{abstract}

\vfill\eject


\section{Introduction}
The nontrivial vacuum structure of $SU(2)$ Yang-Mills-Higgs (YMH) theory
allows for the existence of regular non-perturbative finite mass 
solutions, such as spherically symmetric monopoles \cite{mono}, 
axially symmetric multimonopoles \cite{WeinbergGuth,RebbiRossi,mmono,KKT} and   
monopole-antimonopole pairs \cite{Rueber,mapKK}. 
Recently, more general static equilibrium solutions
have been constructed, 
representing either chains of $m$ alternating monopoles and antimonopoles, 
carrying charge $\pm n$, 
or vortex ring configurations \cite{KKS}.  

The spherically symmetric `t Hooft--Polyakov monopole of unit charge \cite{mono}
is a topologically stable solution of the field equations. 
In the Bogomol'nyi-Prasad-Sommerfield  (BPS) limit 
of vanishing Higgs potential 
axially symmetric multimonopole configurations 
are known analytically \cite{mmono}. In these solutions the nodes of the Higgs 
field are superimposed at a single point. 
In the BPS limit
repulsive and attractive forces between monopoles exactly compensate and 
BPS monopoles experience no net interaction \cite{Manton77}. Indeed,
calculating the number of zero modes  
of the configurations shows that they can be continuously
deformed into systems of individual monopoles with unit topological charge 
\cite{Wein-zero}. 
When the Higgs field becomes massive, 
the fine balance of forces between the monopoles is broken 
since the corresponding attractive 
Yukawa interaction becomes short-ranged, and consequently
the non-BPS monopoles experience repulsion \cite{KKT}. 

As shown by Taubes \cite{Taubes}, each topological sector contains
besides the (multi)monopole solutions further regular, finite mass solutions,
which do not satisfy the first order Bogomol'nyi equations, 
but only the set of second order field equations,
even for vanishing Higgs potential.
Such solutions, representing for instance static axially symmetric 
monopole-antimonopole chain and vortex ring configurations
\cite{KKS},
form saddlepoints of the energy functional,
and possess a mass above the Bogomol'nyi bound. 
They exist because the attractive short-range forces 
between the poles, that are 
mediated by the $A_\mu^3$ vector boson and the Higgs boson, 
are balanced by the repulsion, 
which is mediated by the massive vector bosons $A_\mu^\pm$.

In the topologically trivial sector 
the simplest of these saddlepoint solutions represents a
monopole-antimonopole pair, forming a magnetic dipole \cite{Rueber,mapKK}. 
When the charge $\pm n$ of the monopole and antimonopole 
increases beyond $n=2$, it becomes favourable
for the monopole-antimonopole system to form a vortex ring,
at least for small values of the Higgs boson mass.
Likewise, larger monopole-antimonopole chains then
form several vortex rings \cite{KKS}.
For large values of the Higgs boson mass
also more complicated configurations can appear, which consist of 
monopole-antimonopole pairs or chains as well as vortex rings \cite{KKS}.
The presence of an external interaction
is also known to change the node structure of a configuration \cite{S}.

In the present note we investigate the dependence of
such YMH solutions on the strength of the Higgs selfcoupling $\lambda$,
and thus the value of the Higgs boson mass.
We find that, for large values of the Higgs selfcoupling,
new branches of equilibrium configurations arise
for monopole-antimonopole systems with $n=3$.
In particular, we report the existence of new types of solutions
with winding number $n=3$ and $m=2,3,4$
and compare their properties to those of the known solutions
\cite{KKS}.

In section 2 we recall $SU(2)$ YMH theory, 
and present the axially symmetric Ansatz 
and the boundary conditions. 
We then discuss in section 3 the $\lambda$-dependence
of the new solutions and their properties.

\boldmath
\section{\bf $SU(2)$ Yang-Mills-Higgs solutions}
\unboldmath

\subsection{\bf Action}
We consider $SU(2)$ Yang-Mills-Higgs 
theory with action 
\begin{equation} \label{action}
S =  \int \left\{ - \,\frac{1}{2} {\rm Tr} 
\,\left( F_{\mu\nu}F^{\mu\nu} \right)
-\frac{1}{4} {\rm Tr}
\left(  D_\mu \Phi\, D^\mu \Phi  \right)
-\frac{\lambda}{8} \,
 {\rm Tr} \left[ \left(\Phi^2 - \eta^2 \right)^2 \right]
\right\} d^4 x
\end{equation}
with su(2) gauge potential $A_\mu = A_\mu^a \tau^a/2$,
field strength tensor
$F_{\mu\nu} = \partial_\mu A_\nu - \partial_\nu A_\mu + i e [A_\mu, A_\nu]$,
and covariant derivative of the Higgs field
$D_\mu \Phi = \partial_\mu \Phi +i e [A_\mu, \Phi]$.
$e$ denotes the gauge coupling constant, $\eta$ the vacuum expectation
value of the Higgs field and $\lambda$ the strength of the Higgs selfcoupling.

\subsection{Ansatz}

For the gauge and Higgs field we employ the Ansatz \cite{KKS}
\begin{eqnarray}
A_\mu dx^\mu
& = &
\left( \frac{K_1}{r} dr + (1-K_2)d\theta\right)\frac{\tau_\vphi^{(n)}}{2e}
\nonumber \\
&-& n \sin\theta \left( K_3\frac{\tau_r^{(n,m)}}{2e}
                     +(1-K_4)\frac{\tau_\theta^{(n,m)}}{2e}\right) d\vphi
\ , \label{ansatzA} \\
\Phi
& = &
\eta \left( \Phi_1\tau_r^{(n,m)}+ \Phi_2\tau_\theta^{(n,m)} \right) \  ,
\label{ansatzPhi}
\end{eqnarray}
where the $su(2)$ matrices
$\tau_r^{(n,m)}$, $\tau_\theta^{(n,m)}$, and $\tau_\vphi^{(n)}$
are defined as products of the spatial unit vectors
\begin{eqnarray}
{\hat e}_r^{(n,m)} & = & \left(
\sin(m\theta) \cos(n\vphi), \sin(m\theta)\sin(n\vphi), \cos(m\theta)
\right)\ , \nonumber \\
{\hat e}_\theta^{(n,m)} & = & \left(
\cos(m\theta) \cos(n\vphi), \cos(m\theta)\sin(n\vphi), -\sin(m\theta)
\right)\ , \nonumber \\
{\hat e}_\vphi^{(n)} & = & \left( -\sin(n\vphi), \cos(n\vphi), 0 \right)\ ,
\label{unit_e}
\end{eqnarray}
with the Pauli matrices $\tau^a$. 

The four gauge field functions $K_i$ and two Higgs field functions 
$\Phi_i$ depend on the coordinates $r$ and $\theta$, only.
With this Ansatz the general field equations
reduce to six PDEs in the coordinates $r$ and $\theta$.

The Ansatz possesses a residual U(1) gauge symmetry. 
To fix the gauge we impose
the condition $r\partial_r K_1 - \partial_\theta K_2 = 0$ \cite{KKT}.
We further introduce the dimensionless coordinate 
${\tilde x} = er\eta$ and rescale the 
Higgs field $\tilde \Phi  = \Phi/\eta$.

\subsection{\bf Boundary conditions}

To obtain globally regular solutions with the proper symmetries,
we impose appropriate boundary conditions \cite{KKS}.

{\sl Boundary conditions at the origin}

Regularity of the solutions at the origin ($r=0$) 
requires the conditions
\begin{equation}
K_1(0,\theta)= K_3(0,\theta)= 0\ , \ \ \ \
K_2(0,\theta)= K_4(0,\theta)= 1 \ ,
\end{equation}
\begin{equation}
\sin(m\theta) \Phi_1(0,\theta) + \cos(m\theta) \Phi_2(0,\theta) = 0 \ ,
\end{equation}
\begin{equation}
\left.\partial_r\left[\cos(m\theta) \Phi_1(r,\theta)
              - \sin(m\theta) \Phi_2(r,\theta)\right] \right|_{r=0} = 0 \ ,
\end{equation}
i.e.~$\Phi_\rho(0,\theta) =0$, $\partial_r \Phi_z(0,\theta) =0$.

{\sl Boundary conditions at infinity}

At infinity we require that solutions in the vacuum sector tend to
a gauge transformed trivial solution,
$$
\Phi \ \longrightarrow \eta U \tau_z U^\dagger \   , \ \ \
A_\mu \ \longrightarrow  \ \frac{i}{e} (\partial_\mu U) U^\dagger \ ,
$$
and that solutions in the sector with topological charge $n$
tend to
$$
\Phi  \longrightarrow  U \Phi_\infty^{(1,n)} U^\dagger \   , \ \ \
A_\mu \ \longrightarrow \ U A_{\mu \infty}^{(1,n)} U^\dagger
+\frac{i}{e} (\partial_\mu U) U^\dagger \  ,
$$
where
$$ \Phi_\infty^{(1,n)} =\eta \tau_r^{(1,n)}\ , \ \ \
A_{\mu \infty}^{(1,n)}dx^\mu =
\frac{\tau_\vphi^{(n)}}{2e} d\theta
- n\sin\theta \frac{\tau_\theta^{(1,n)}}{2e} d\vphi
$$
is the asymptotic solution of a charge $n$ multimonopole,
and $U = \exp\{-i k \theta\tau_\vphi^{(n)}\}$.

In terms of the functions $K_1 - K_4$, $\Phi_1$, $\Phi_2$ these boundary
conditions read
\begin{equation}
K_1 \longrightarrow 0 \ , \ \ \ \
K_2 \longrightarrow 1 - m \ , \ \ \ \
\label{K12infty}
\end{equation}
\begin{equation}
K_3 \longrightarrow \frac{\cos\theta - \cos(m\theta)}{\sin\theta}
\ \ \ m \ {\rm odd} \ , \ \ \
K_3 \longrightarrow \frac{1 - \cos(m\theta)}{\sin\theta}
\ \ \ m \ {\rm even} \ , \ \ \
\label{K3infty}
\end{equation}
\begin{equation}
K_4 \longrightarrow 1- \frac{\sin(m\theta)}{\sin\theta} \ ,
\label{K4infty}
\end{equation}
\begin{equation}        \label{Phiinfty}
\Phi_1\longrightarrow  1 \ , \ \ \ \ \Phi_2 \longrightarrow 0 \ .
\end{equation}

{\sl Boundary conditions along the symmetry axis}

The boundary conditions along the $z$-axis
($\theta=0$ and $\theta=\pi $) are determined by the
symmetries,
\begin{equation}
K_1 = K_3 = \Phi_2 =0 \ , \ \ \  \
\partial_\theta K_2 = \partial_\theta K_4 = \partial_\theta \Phi_1 =0 \ .
\end{equation}


\section{Numerical results}

The numerical calculations are performed with help of the package FIDISOL,
based on the Newton-Raphson iterative procedure \cite{FIDI}.
We solve the system of 6 coupled non-linear partial differential equation
numerically, subject to the above set of boundary conditions, 
employing the compact radial coordinate 
$\bar x={\tilde x}/(1+{\tilde x}) \in [0:1]$. 

We mainly present results for the systems with $n=3$ and $m=2,~3,~4$,
where all quantities shown are dimensionless.
In particular, we illustrate the dependence 
of the structure of these systems of solutions
on the strength of the Higgs selfcoupling $\lambda$. 

In the limit of vanishing and small Higgs selfcoupling,
these $n=3$ solutions have been studied before \cite{KKS}.
When $m=2$, 
they consist of a single vortex ring in the $xy$-plane.
When $m=3$, they consist of
two opposite vortex rings located symmetrically above and below
the $xy$-plane together with a triple pole at the origin.
When $m=4$, they consist of
two like vortex rings located symmetrically above and below
the $xy$-plane.

As $\lambda$ is increased from zero, each of these solutions 
gives rise to a branch of solutions, to which
we refer as the respective fundamental branches.
Interestingly, at critical values of $\lambda$, pairs of
new branches of solutions appear, whose node structure
differs from the node structure of the solutions of 
the corresponding fundamental branches.

\boldmath
\subsection{Topologically trivial sector: $m=2$ and $m=4$}
\unboldmath

{\sl $m=2$}:

Let us start with the simplest case, the $n=3$, $m=2$ solutions,
which possess a single vortex ring along their fundamental branch.
As $\lambda$ increases, the mass of the solutions
increases along the fundamental branch.
At the same time, the radius of the single dipole ring 
in the $xy$-plane decreases slowly.
The $\lambda$-dependence of the mass 
and the location of the vortex ring $\rho_0$
of the solutions along the fundamental branch are shown in 
Fig.~\ref{f-1}.
\begin{figure}
\lbfig{f-1}
\begin{center}
\includegraphics[height=.25\textheight, angle =0]{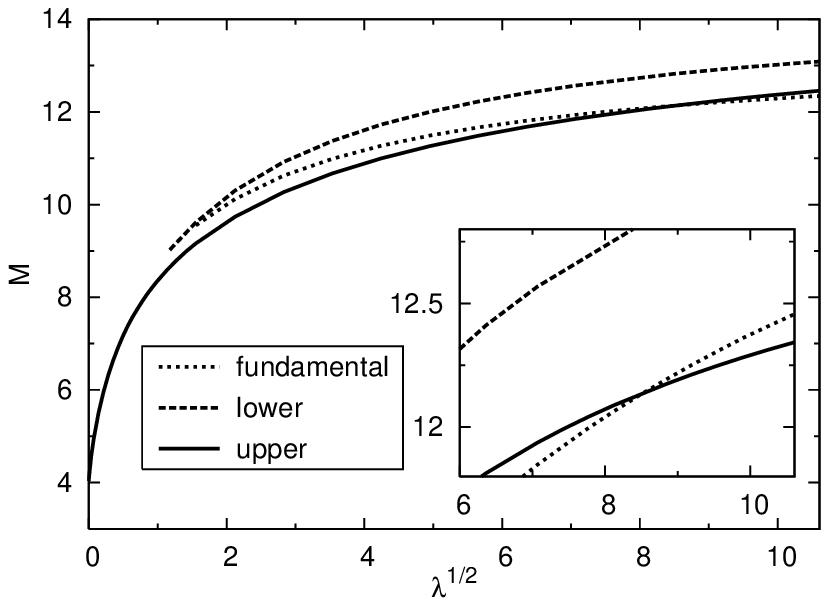}
\includegraphics[height=.25\textheight, angle =0]{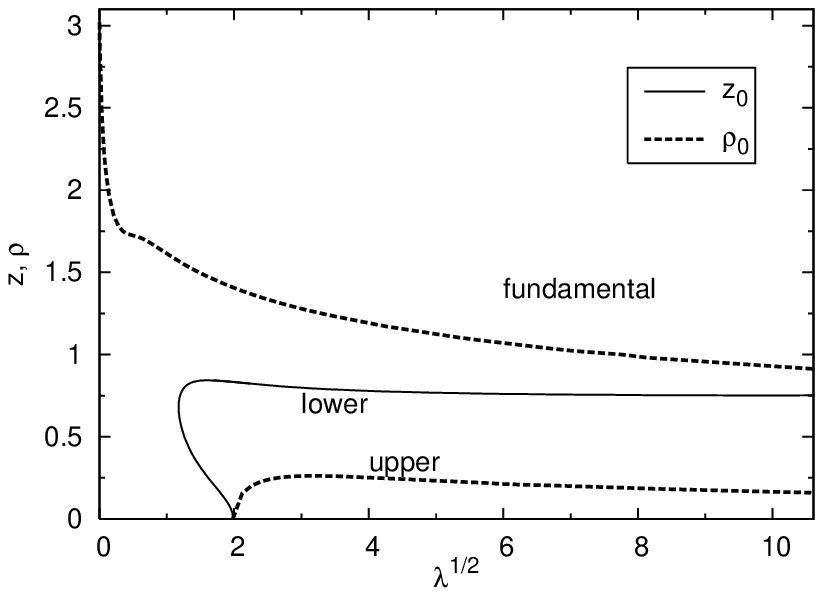}
\end{center}
\caption{left: 
The mass of the fundamental branch as well as of the new
lower (mass) and upper (mass) branch of $n=3$, $m=2$ solutions 
versus the Higgs selfcoupling $\lambda$.
right: 
The location of the nodes of the Higgs field for the same 
set of solutions.
($\rho_0$ denotes the radius of the rings in the $xy$-plane, $z_0$
the location of the isolated nodes on the symmetry axis.)
}
\end{figure}

While the fundamental branch persists as $\lambda$ increases,
a new solution appears at a critical value $\lambda_c^1=1.382$.
This solution has higher mass than the fundamental solution,
and it has a different node structure:
its Higgs field possesses two isolated nodes on the symmetry axis,
representing a monopole-antimonopole pair with charges $\pm 3$.

As $\lambda$ is increased now, two new branches of solutions arise
from this critical solution, which differ in mass.
The solutions on the lower (mass) branch retain the node structure of the
critical solution. 
Their two isolated nodes on the symmetry axis
change only slightly in distance with increasing $\lambda$. 
Their energy density exhibits two tori,
whose position is associated with the nodes of the Higgs field,
as seen in Fig.~\ref{f-2}.
\begin{figure}
\lbfig{f-2}
\begin{center}
\includegraphics[height=.35\textheight, angle =-90]{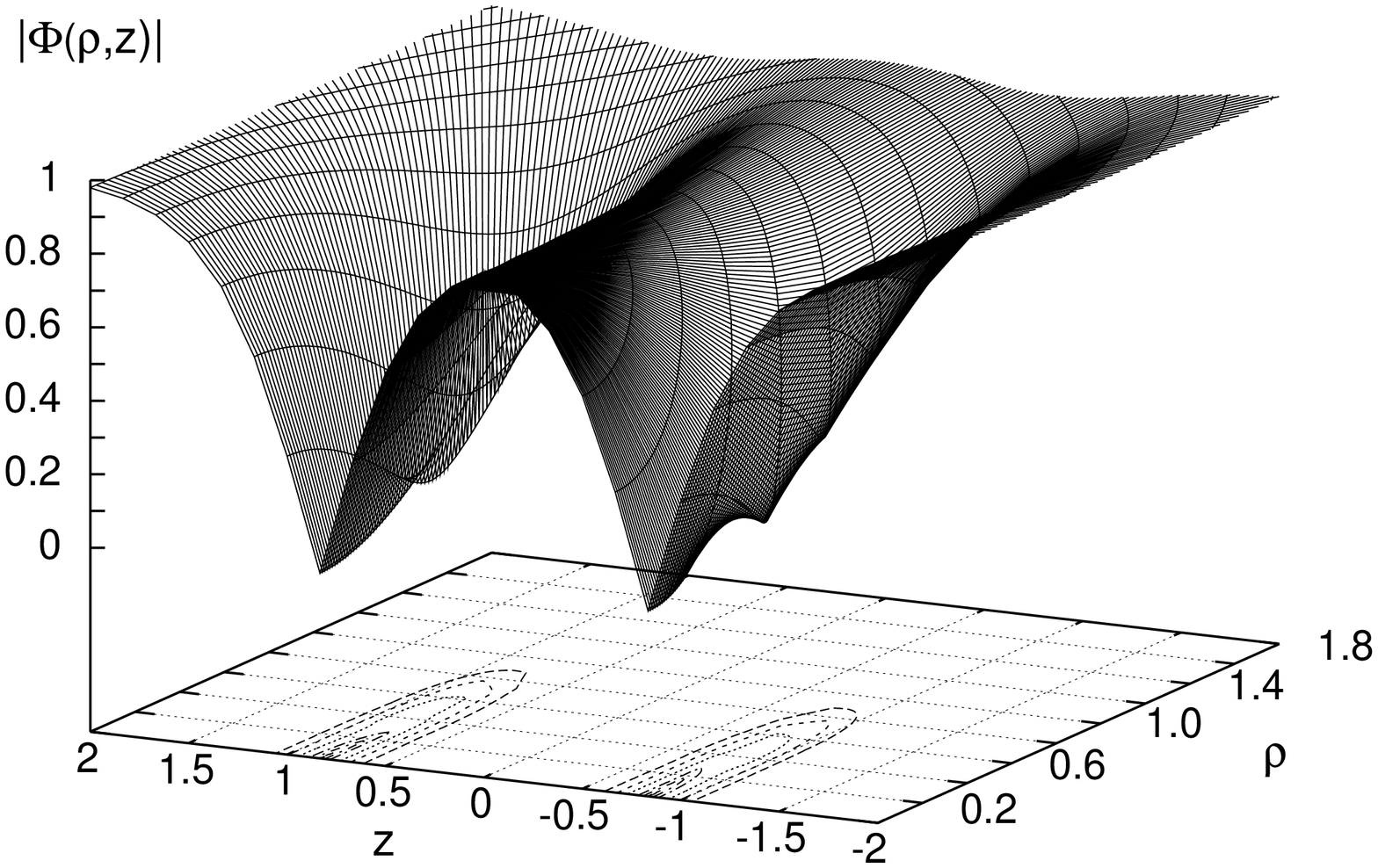}
\includegraphics[height=.35\textheight, angle =-90 ]{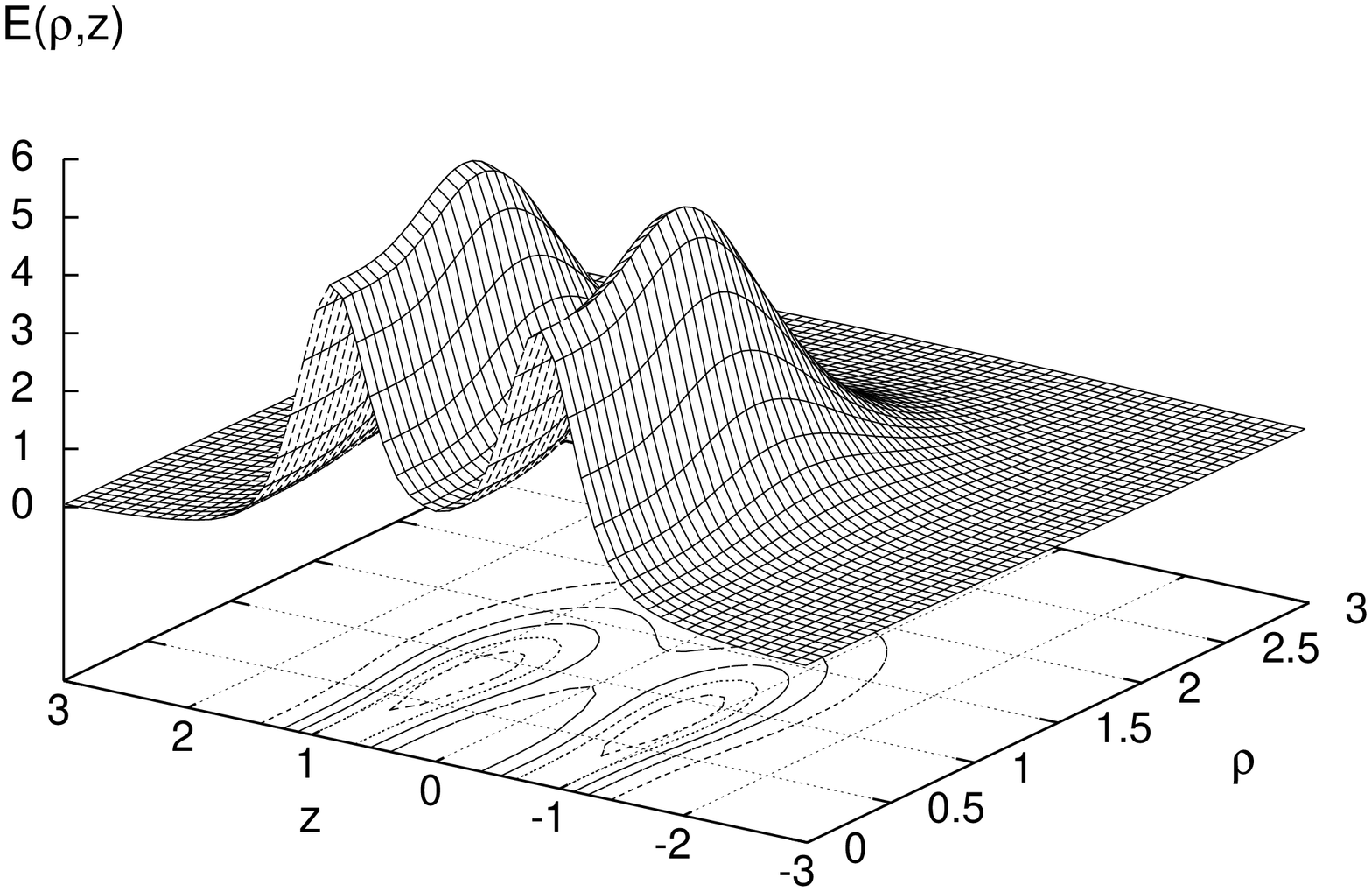}
\end{center}
\caption{
The modulus of the Higgs field (left) and the energy density (right) 
of the $n=3$, $m=2$ solution on the lower branch
are shown as functions of the coordinates $z$ and $\rho$ 
for $\lambda = 2.0 $.
}
\end{figure}

The solutions on the upper (mass) branch, in contrast,
do not retain the node structure of the
critical solution for long.
Their isolated nodes on the symmetry axis approach
each other rapidly, and merge at the origin
at a second critical value $\lambda_c^2=3.941$.
Thus at $\lambda_c^2$ we observe a transition
from a monopole-antimonopole pair solution to a vortex ring solution.
Beyond $\lambda_c^2$
the radius of the ring first increases rapidly
and then decreases slowly again.
Both new branches of solutions are also shown in Fig.~\ref{f-1}.

We illustrate the new solutions further in Fig.~\ref{f-3},
where we exhibit the modulus of the
Higgs field $|\Phi|$ and the gauge function $K_2$
along the symmetry axis and in the $xy$-plane
for two values of coupling constant $\lambda$.
For $\lambda =2$ both solutions possess
isolated nodes on the symmetry axis,
but for the solution on the lower (mass) branch
these are farther apart.
In contrast,
for $\lambda =112.5$ the solution on the upper branch has a small ring
in the $xy$-plane.
\begin{figure}
\lbfig{f-3}
\begin{center}
\includegraphics[height=.25\textheight, angle =0 ]{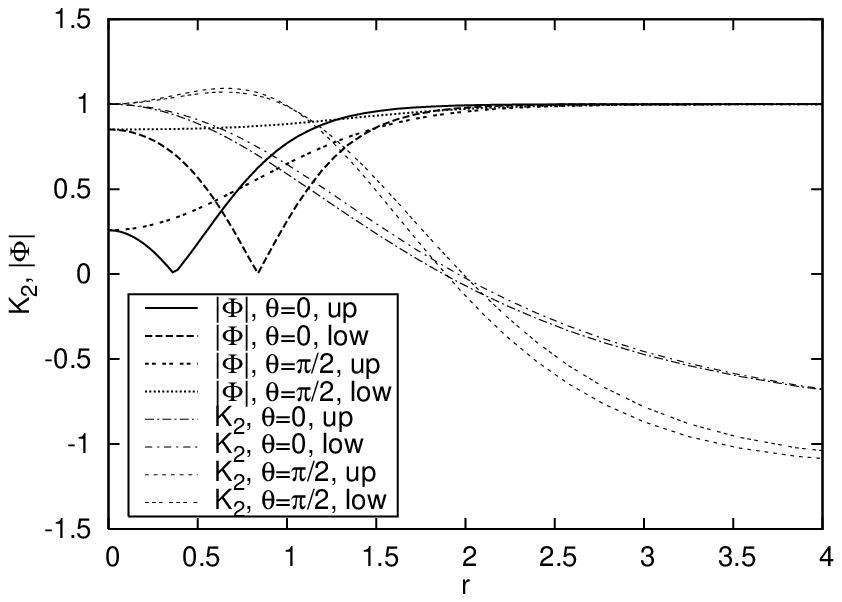}
\includegraphics[height=.25\textheight, angle =0 ]{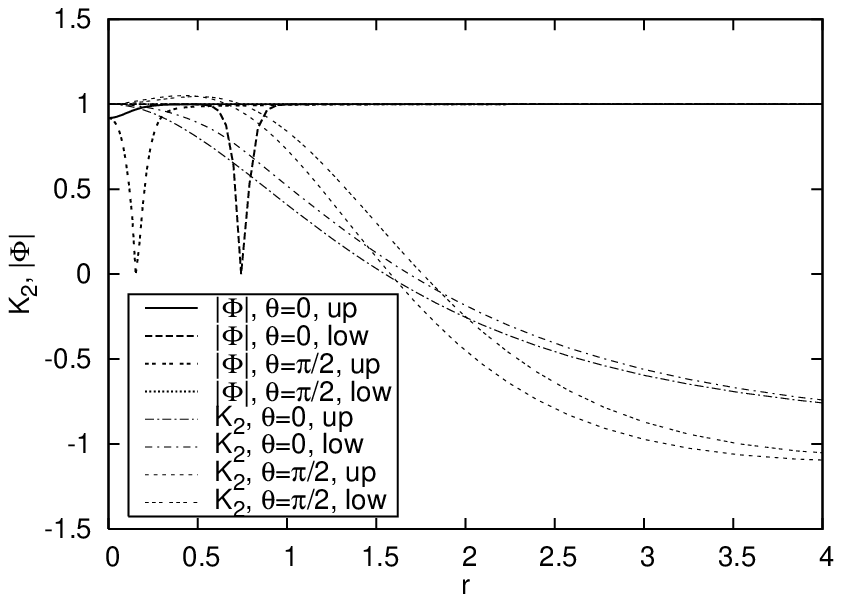}
\end{center}
\caption{
The modulus of the Higgs field $\Phi$ and the gauge field function $K_2$
of $n=3$, $m=2$ solutions on the lower and upper branch
are shown as functions of the radial coordinate $r$ 
along the $z$-axis ($\theta=0$) and in the $xy$-plane ($\theta=\pi/2$)
for $\lambda = 2$ (left) and $\lambda = 112.5$ (right).
}
\end{figure}

Although beyond $\lambda_c^2$ the new upper branch solutions possess
the same node structure as the solutions on the fundamental branch,
the size of their vortex ring is much smaller
and their mass remains considerably higher.
However, since the mass of the new lower branch increases
more slowly with $\lambda$ than the mass of the fundamental branch,
a further critical value of $\lambda$ appears,
$\lambda_c^3 \approx 72.8$, where the mass of the solution of the fundamental branch
coincides with the mass of the solution of the lower branch.
$\lambda_c^3$ thus marks the transition, where it becomes energetically
favourable for the field configuration to have two triple nodes
on the symmetry axis instead of a single large
vortex ring in the $xy$-plane.

For larger values of the Higgs selfcoupling
the subtle interplay between repulsive and attractive forces thus allows for  
more than one non-trivial equilibrium configuration. 
Analyzing the various contributions to the total mass of these
configurations shows, that the kinetic energy of the Higgs field is 
smallest for the solutions on the fundamental branch
(except for a small range of $\lambda$ close
to the first critical point).
But the potential energy of the Higgs field
and the kinetic energy of the gauge fields 
are smallest for the solutions on the new lower branch,
for larger values of $\lambda$.
Concerning the total energy balance it then becomes favourable
for the Higgs field to form pointlike isolated nodes
instead of extended vortex-like nodes,
thus causing a transition
from a vortex ring to a monopole-antimonopole pair configuration
at a critical value of $\lambda$.


\vspace{0.5cm}
\noindent
{\sl $m=4$}:

We now turn to the $n=3$, $m=4$ solutions.
The fundamental $n=3$, $m=4$ solutions possess two vortex rings
located symmetrically with respect to the $xy$-plane.
Their radius $\rho_1$ almost coincides with their distance
$z_1$ from the plane, as seen in Fig.~\ref{f-4}.
Their location and size varies slowly with $\lambda$.
\begin{figure}
\lbfig{f-4}
\begin{center}
\includegraphics[height=.25\textheight, angle =0]{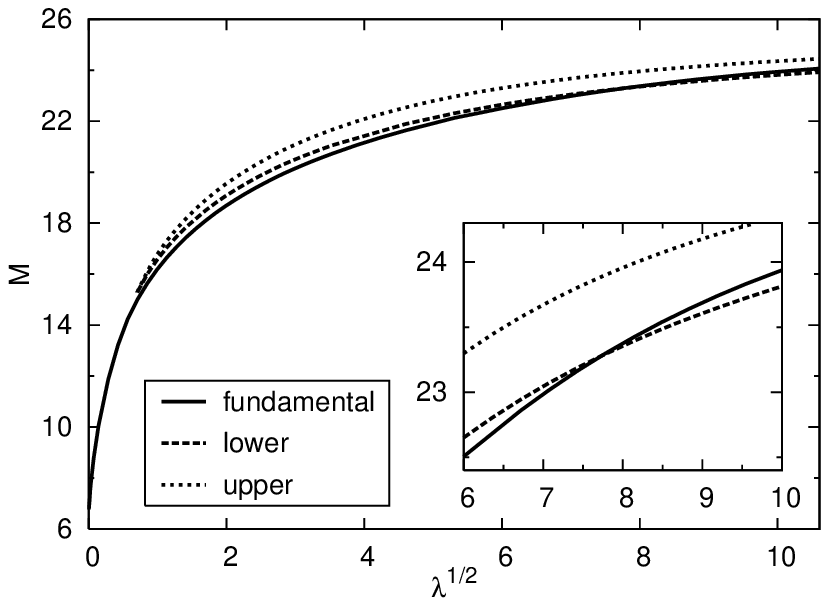}
\includegraphics[height=.25\textheight, angle =0]{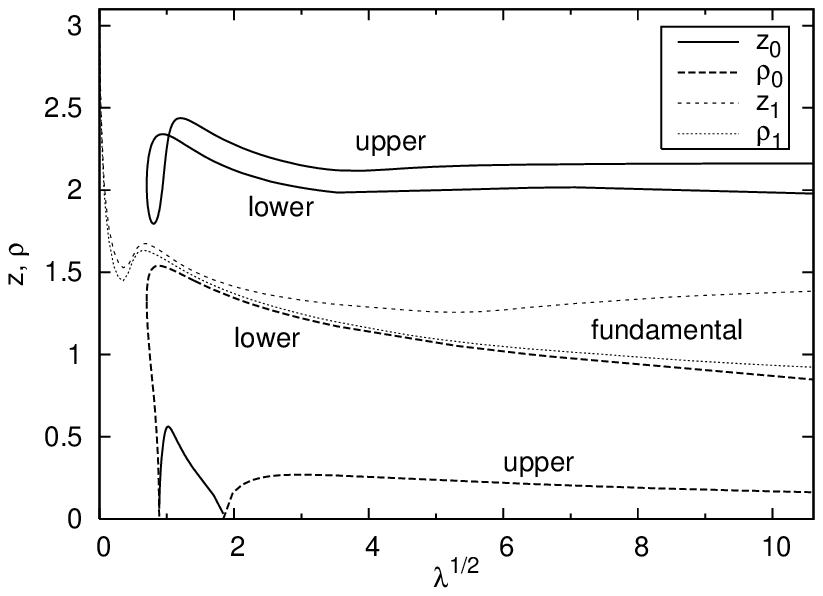}
\end{center}
\caption{left:
The mass of the fundamental branch as well as of the new
lower (mass) and upper (mass) branch of $n=3$, $m=4$ solutions
versus the Higgs selfcoupling $\lambda$.
right:
The location of the nodes of the Higgs field for the same
set of solutions.
($\rho_0$ denotes the radius of the rings in the $xy$-plane, $z_0$
the location of the isolated nodes on the symmetry axis,
$\rho_1$ and $z_1$ denote the location of the rings above the $xy$-plane.)
}
\end{figure}

As $\lambda$ is increased, again a critical value $\lambda_c^1=0.491$
is encountered, 
where two new branches of solutions arise, possessing 
higher mass and a different node structure than the solutions
on the fundamental branch.
The new solutions possess two outer nodes on the symmetry axis,
as well as a vortex ring in the $xy$-plane.
Thus these solutions present a new type of solution
with mixed node structure.

With increasing $\lambda$ the solutions on
the lower (mass) branch again retain this node structure,
keeping two isolated nodes on the symmetry axis
and a vortex ring in the $xy$-plane.
The solutions on the upper (mass) branch, however,
again do not retain this node structure for long. 
Their single vortex ring in the $xy$-plane decreases rapidly in size, 
and reaches zero size at a second critical value $\lambda_c^2=0.786$.
At $\lambda_c^2$
we then observe the transition to a monopole-antimonopole
chain solution, possessing four isolated nodes on the symmetry axis.
Their node structure is illustrated in Fig.~\ref{f-5},
where the modulus of the Higgs field is exhibited for both (types of)
new solutions at $\lambda = 1$.
\begin{figure}
\lbfig{f-5}
\begin{center}
\includegraphics[height=.35\textheight, angle =-90]{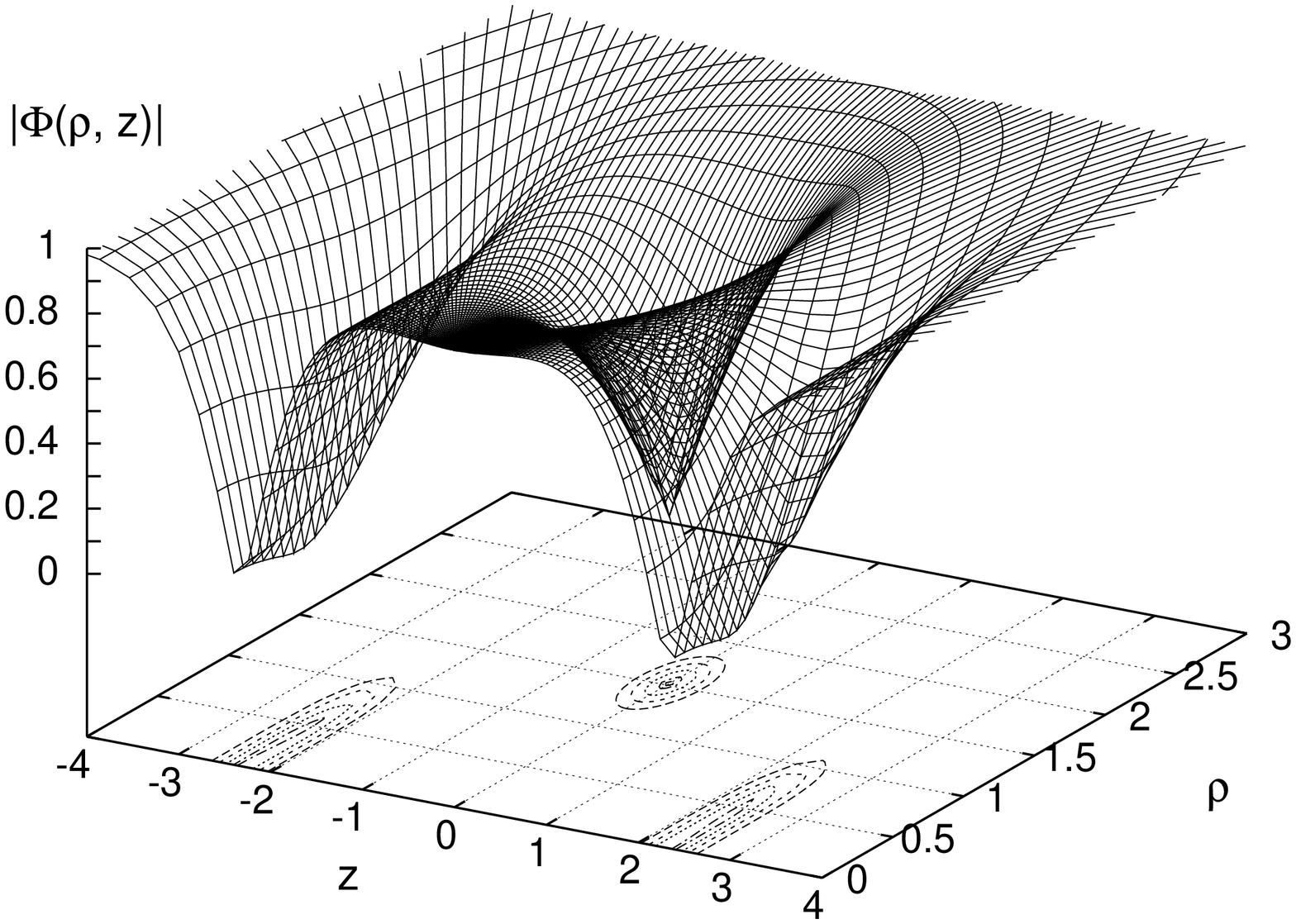}
\includegraphics[height=.35\textheight, angle =-90]{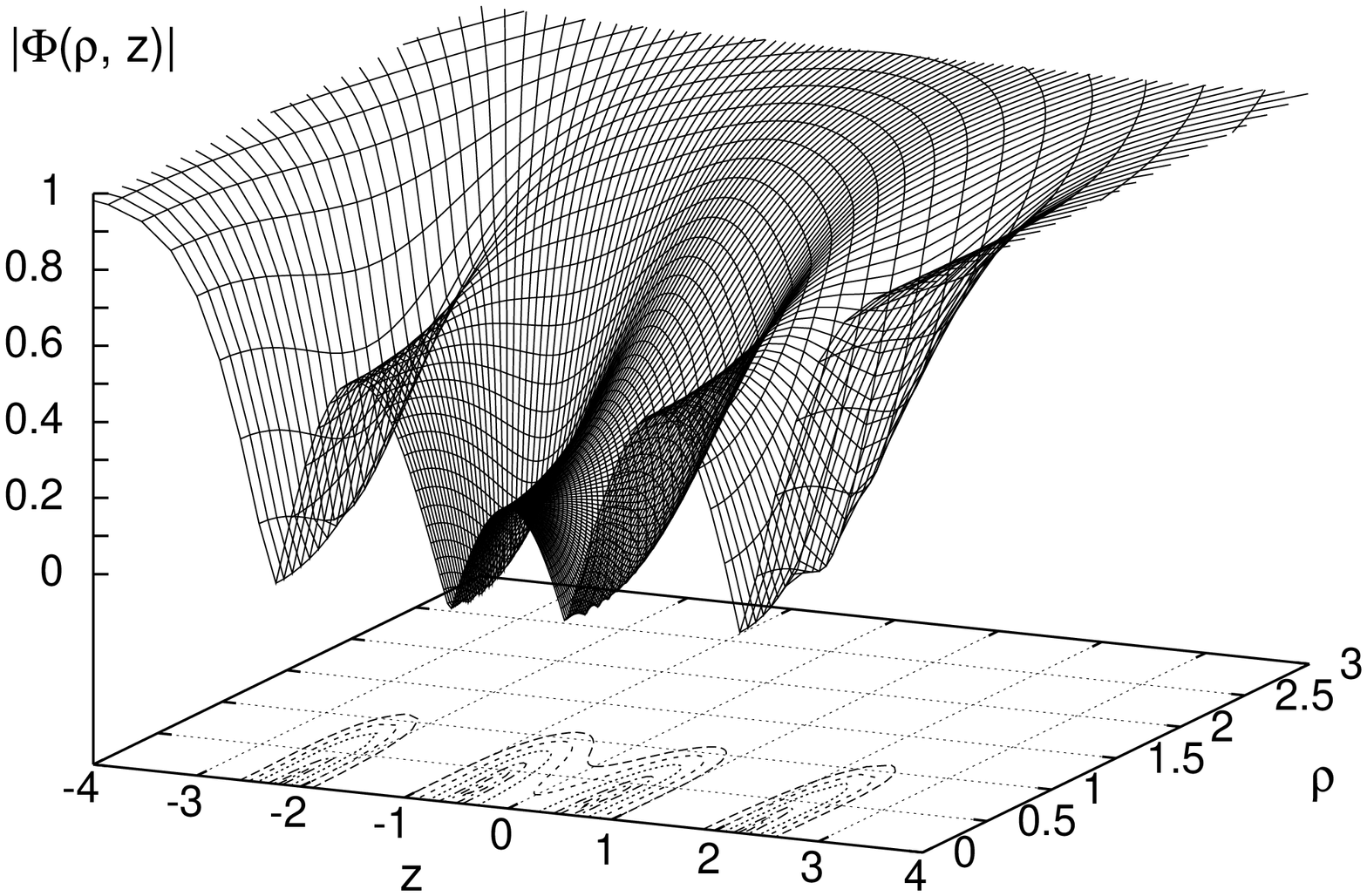}
\end{center}
\caption{
The modulus of the Higgs field of the lower branch (right)
and upper branch (left) 
$n=3$, $m=4$ solutions
is shown as a function of the coordinates $z$ and $\rho$ for $\lambda = 1$.
}
\end{figure}

Interestingly, the new inner nodes approach each other again,
and coalesce at the origin at a further critical value, $\lambda_c^3=3.406$.
Beyond $\lambda_c^3$ the solutions possess again
two outer nodes on the symmetry axis,
and a small vortex ring in the $xy$-plane.
The mass and the nodes along the new branches of solutions 
are also shown in Fig.~\ref{f-4}.

Concerning the mass of the new solutions,
we again observe a transition
between the fundamental branch and the new lower (mass) branch:
beyond $\lambda_c^4 \approx 59.8$ the new lower branch solutions
are energetically favourable, i.e.~it becomes again
advantageous to exchange vortex rings for isolated nodes.
The new lowest mass solution thus contains instead of
two vortex rings only a single vortex ring and two nodes on
the symmetry axis beyond $\lambda_c^4$.

Continuing this reasoning it is tempting to conjecture, 
that a further critical value of $\lambda$ might exist, 
where another pair of branches would appear with now
four isolated nodes on the symmetry axis,
representing thus monopole-antimonopole chains,
and the solutions on this (conjectured) lower (mass) branch would 
become the energetically most favourable configurations
for high values of $\lambda$.

\boldmath
\subsection{Topologically nontrivial sector: $m=3$}
\unboldmath

So far we considered solutions of the topologically trivial sector.  
Let us now address the $\lambda$-dependence of solutions in the sector 
with topological charge $n=3$. 
For $\lambda \rightarrow 0$,
the $n=3$, $m=3$ solutions possess a triply charged monopole at the origin
and two oppositely oriented vortex rings located symmetrically
above and below the $xy$-plane \cite{KKS}.

Based on the observations in the topologically trivial sector
we expect a bifurcation at a critical value of $\lambda$,
where two new branches of solutions appear, which possess
a node structure different from the solutions on the
fundamental branch.
Furthermore, for high values of $\lambda$
the energetically most favourable solutions
should represent monopole-antimonopole chains.

Constructing the solutions confirms these expectations,
but in a surprising way:
the node structure of the solutions changes already
along the fundamental branch, 
and the fundamental branch and the new lower branch merge and end
at a critical value of $\lambda$,
while beyond this critical value only the new upper branch persists,
as illustrated in Fig.~\ref{f-6}.
\begin{figure}
\lbfig{f-6}
\begin{center}
\includegraphics[height=.26\textheight, angle =0]{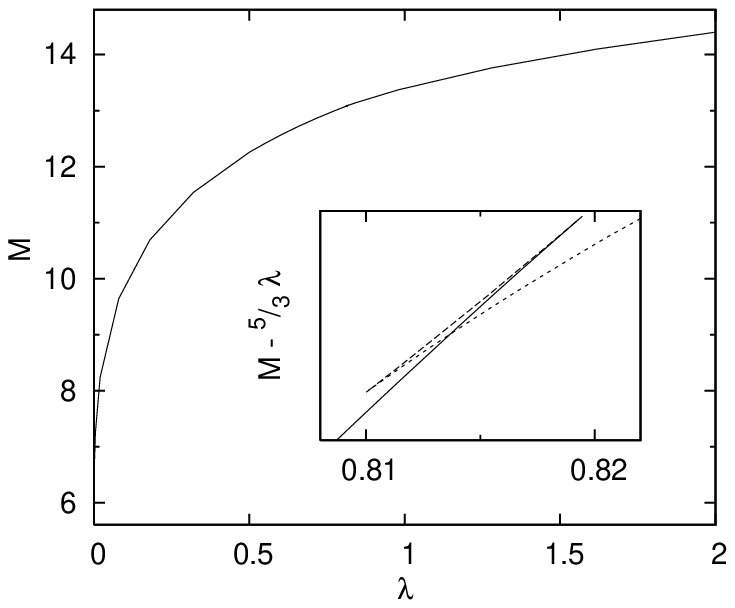}
\includegraphics[height=.26\textheight, angle =0]{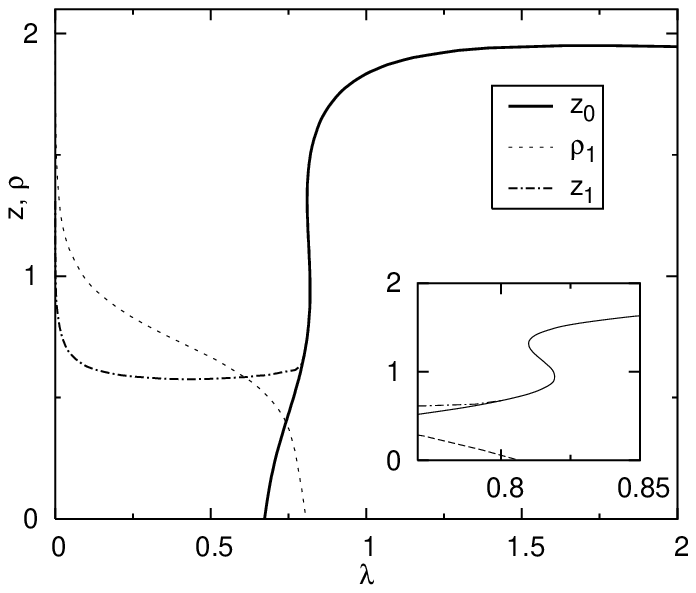}
\end{center}
\caption{
left:
The mass of the fundamental branch as well as of the new
lower (mass) and upper (mass) branch of $n=3$, $m=3$ solutions
versus the Higgs selfcoupling $\lambda$.
right:
The location of the nodes of the Higgs field for the same
set of solutions.
($z_0$ denotes the location of the isolated nodes on the symmetry axis,
$\rho_1$ and $z_1$ denote the location of the rings above the $xy$-plane.)
}
\end{figure}

Considering the $\lambda$-dependence of the solutions in detail,
we observe that with increasing $\lambda$
the radius of the vortex rings decreases
while the vortex rings first move closer towards each other
and then roughly retain their distance.
Then, at a critical value $\lambda_c^1 = 0.673$,
two nodes emerge from the origin and separate from
each other along the $z$-axis \footnote{
These new nodes appear to be encircled by tiny 
rings not exhibited in Fig.~\ref{f-6}.}.
Thus beyond $\lambda_c^1$
the solutions on the fundamental branch possess
three nodes on the symmetry axis and two vortex rings
located symmetrically above and below the $xy$-plane.
As $\lambda$ increases further, 
the new nodes move further apart, 
while the vortex rings shrink to zero size
and merge with the new nodes on the $z$-axis 
at a critical value $\lambda_c^2=0.807$. 
Beyond $\lambda_c^2$, the solutions possess only three isolated 
nodes on the symmetry axis and represent monopole-antimonopole
chains \footnote{
The $\lambda$-dependence of the nodes 
of the solutions with fixed $n=3$ and $m=3$
is very similar to the $n$-dependence
of the nodes of the solutions with fixed $\lambda=0$ and $m=3$
\cite{KKS}.}.

We note though, that for a small range of $\lambda$,
$0.810 \le\lambda\le 0.819$,
three branches of solutions are present,
as seen in Fig.~\ref{f-6}.
Clearly, at $\lambda_c^3=0.810$ two new branches of
solutions appear which possess the node structure of 
monopole-antimonopole chains.
The new lower branch then merges with the
fundamental branch at $\lambda_c^4=0.819$, 
where both branches end,
while the upper branch extends to high values of $\lambda$.
The modulus of the Higgs field and the energy density
of several $n=3$, $m=3$ solutions are illustrated in
Fig.~\ref{f-7}.
\begin{figure}
\lbfig{f-7}
\begin{center}
\includegraphics[height=.26\textheight, angle =0]{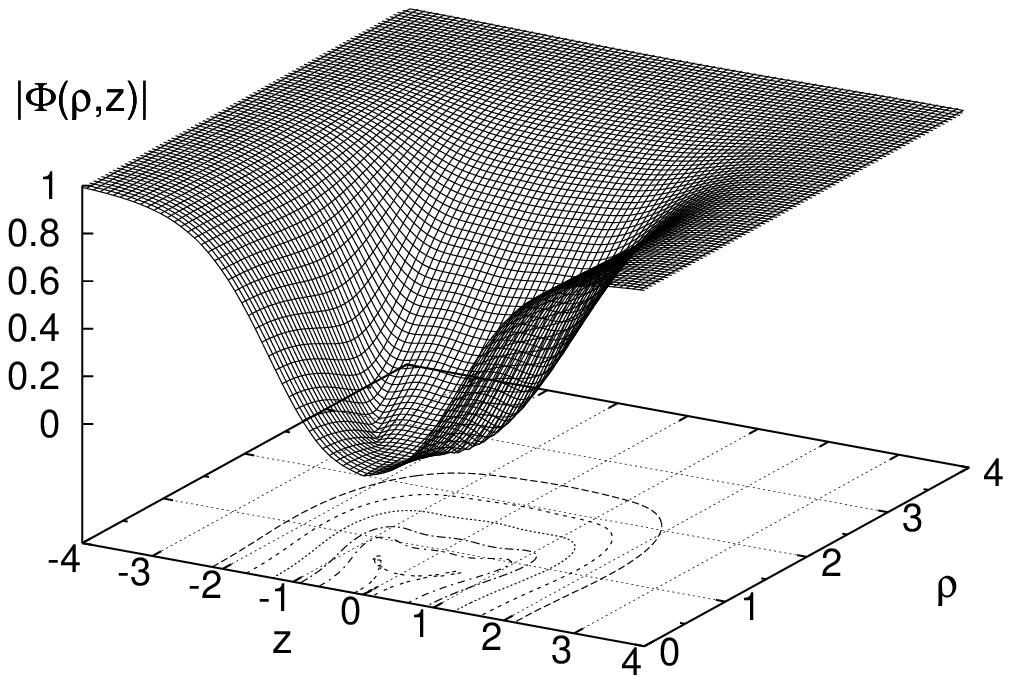}
\includegraphics[height=.26\textheight, angle =0]{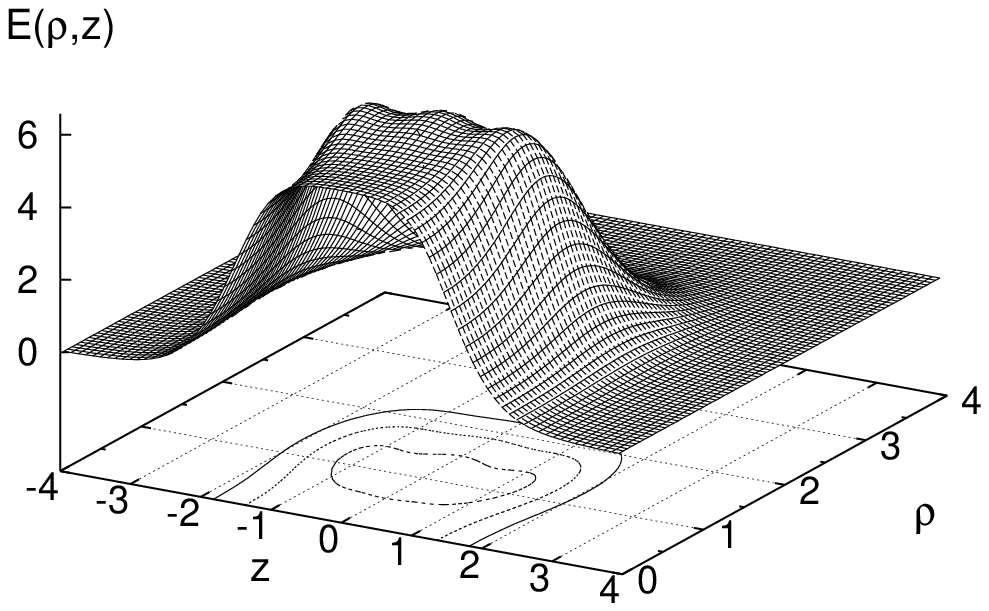}
\includegraphics[height=.26\textheight , angle =0 ]{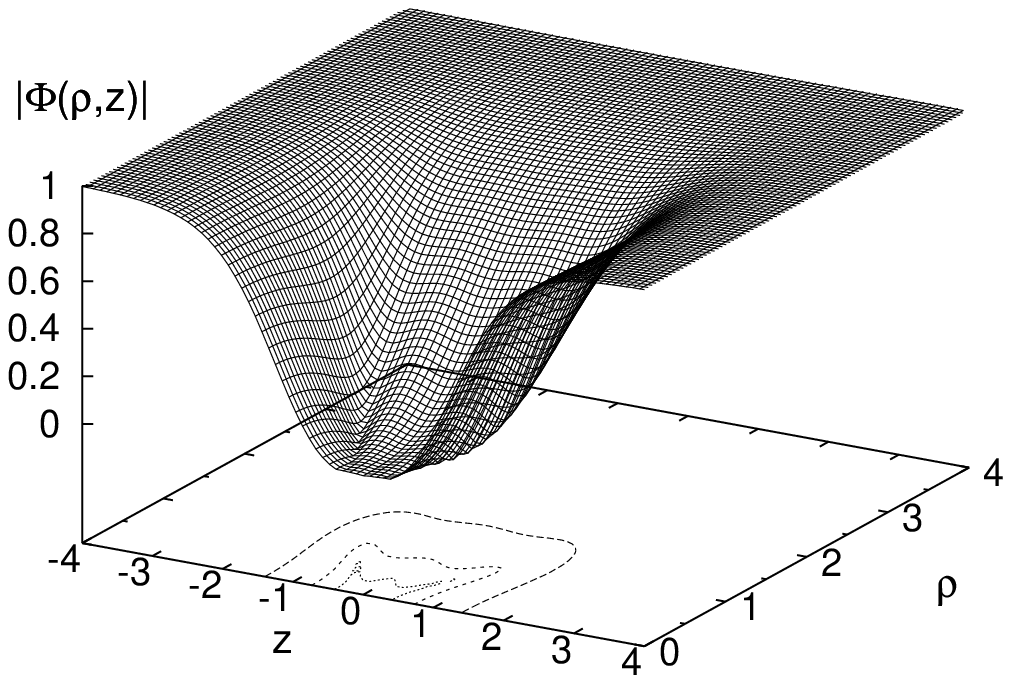}
\includegraphics[height=.26\textheight , angle =0 ]{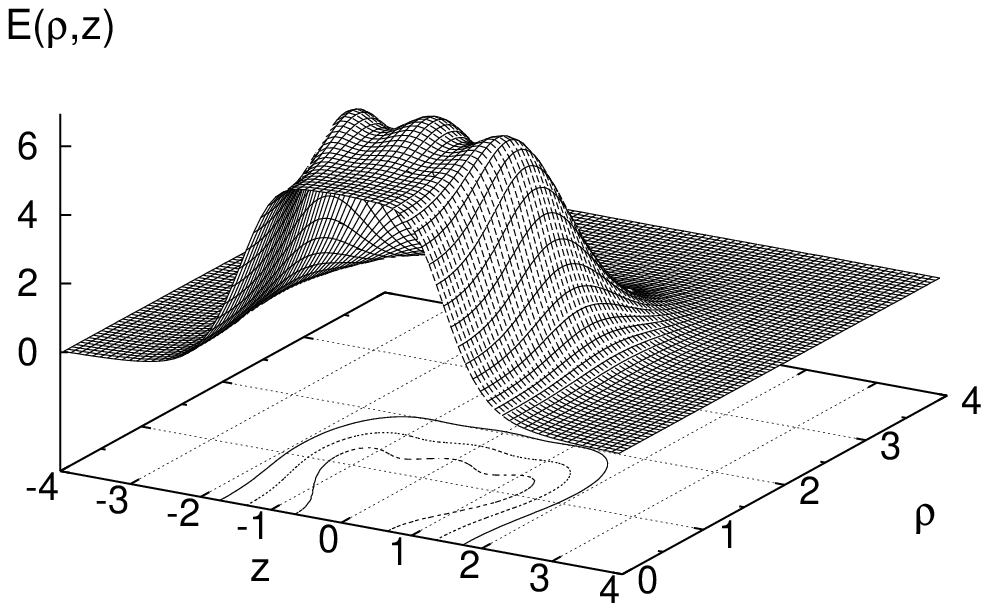}
\includegraphics[height=.26\textheight , angle =0 ]{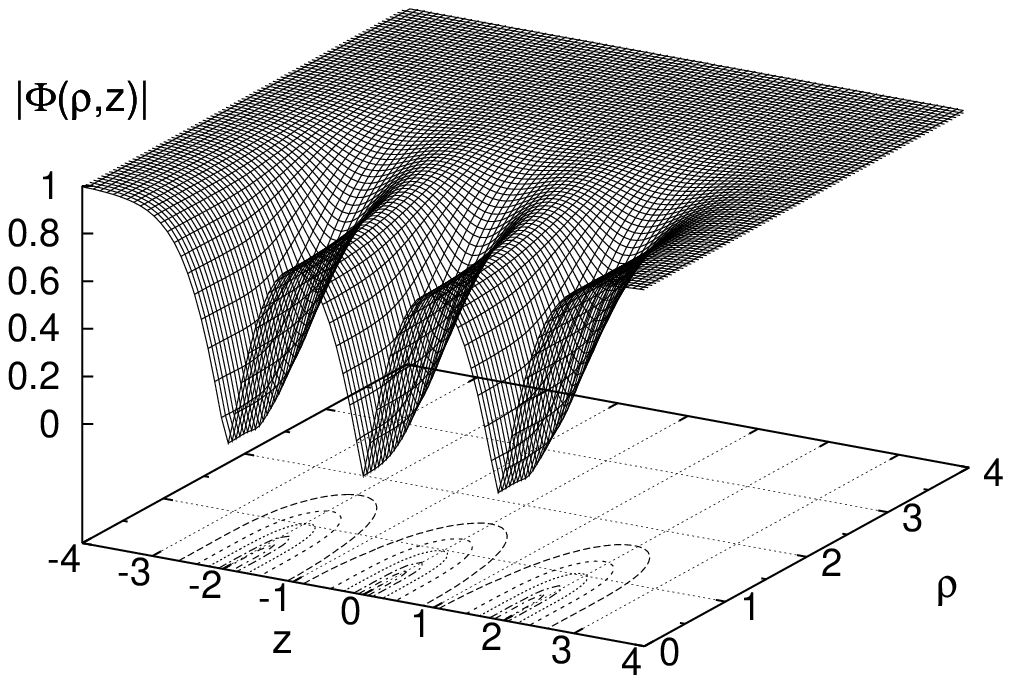}
\includegraphics[height=.26\textheight , angle =0 ]{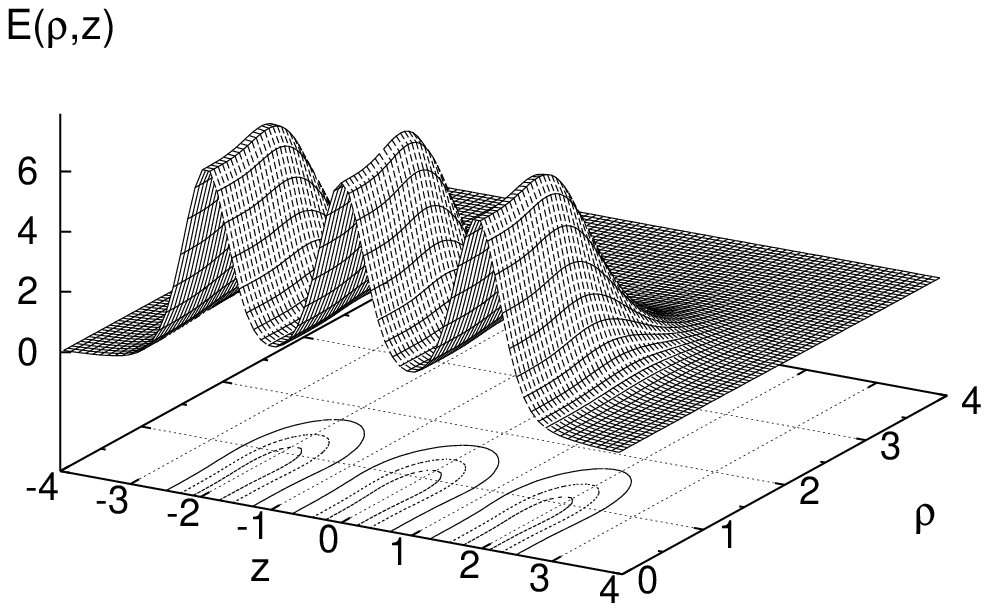}
\end{center}
\caption{
The modulus of the Higgs field (left) and the energy density (right)
of the $n=3$, $m=3$ solutions
are shown as functions of the coordinates $z$ and $\rho$
for $\lambda=0.5$, $\lambda=0.72$ and $\lambda=1.28$.}
\end{figure}

\section{Conclusions}

We have investigated static axially symmetric solutions of the
$SU(2)$ Yang-Mills-Higgs theory,
representing monopole-antimonopole chains and vortex rings,
and obtained new types of solutions, representing
mixed chain-vortex ring configurations.

Starting from vortex ring solutions in the limit of
vanishing Higgs selfcoupling constant $\lambda$,
we observe that at critical values of $\lambda$
pairs of new branches of solutions appear.
Thus for larger values of $\lambda$
the subtle interplay between repulsive and attractive forces 
allows for more than one non-trivial equilibrium configuration
of these systems.

The new branches of solutions possess a different node structure,
where, in particular, vortex rings are replaced by 
isolated nodes on the symmetry axis.
For high values of $\lambda$
these new solutions have the lowest mass.

While we have studied here in detail 
only the systems with $n=3$ and $m=2,~3,~4$,
we conjecture, that this phenomenon is not restricted
to these particular systems but that it is of a more general
nature, implicating an enormous richness of configuration space
for high values of $\lambda$.

Finally, it appears interesting to consider the effects of gravity
on these new types of solutions, and thus to obtain the gravitating analoga
of these regular solutions as well as the corresponding 
non-Abelian black holes solutions, if they exist
\cite{gmono,HKK,MAP,IKK}.

\vspace{0.5cm}
{\bf Acknowledgement}

We would like to acknowledge valuable discussions with Burkhard Kleihaus,
Stephane Nonnemacher, Eugen Radu and Tigran Tchrakian.

\end{document}